\documentclass{iopjournal}

\usepackage{booktabs}
\usepackage{multirow}
\usepackage{siunitx}
\usepackage{amsmath}
\usepackage{amssymb}
\usepackage{hyperref}
\usepackage{xcolor}

\begin{document}

\articletype{Benchmark}

\title{Benchmarking Universal Machine Learning Force Fields for Molecular
Dynamics of Lunar Regolith Minerals}

\author{Ziyu Huang$^{1,*}$\orcid{0000-0002-8624-1264},
        Ken-ichi Nomura$^{2,*}$\orcid{0000-0002-1821-5689} 
        % and
        % Thomas M. Orlando$^{3,4}$\orcid{0000-0002-2422-4506}
        }

\affil{$^1$Daniel Guggenheim School of Aerospace Engineering, Georgia Institute of Technology, 620 Cherry Street, Atlanta, GA 30332, USA 

$^2$Collaboratory for Advanced Computing
and Simulation, University of Southern California, Los
Angeles, CA 90089, USA. 
% $^3$ School of Chemistry and Biochemistry, Georgia Institute of Technology, 620 Cherry Street, Atlanta, GA 30332, USA $^4$School of Physics, Georgia Institute of Technology, 620 Cherry Street, Atlanta, GA 30332, USA
}

\affil{$^*$Author to whom any correspondence should be addressed.}

\email{zyuhuang@gatech.edu, knomura@usc.edu}


\begin{abstract}

Universal machine-learning interatomic potentials provide a promising route for accelerating molecular dynamics simulations of materials, but their transferability to lunar regolith-relevant silicates, oxides, and hydrogen-bearing surface species remains elucidated. Here, we benchmark six foundation models, MACE-MH, MatterSim, SevenNet-0, UPET, UMA, and NequIP-OAM-L, using NVT molecular dynamics simulations of four representative lunar minerals: forsterite, fayalite, ilmenite, and anorthite. Structural fidelity is evaluated using temperature stability, bond-distance statistics, bond-angle distributions, and partial radial distribution functions, with comparison to crystallographic reference data. 
The models reproduce Si--O, Mg--O, Al--O, and Ca--O local environments reasonably well, while Fe--O and Ti--O coordination environments show broader distributions and larger short-timescale fluctuations, highlighting the need for further validation and fine tuning with additional ground truth data for Fe- and Ti-bearing lunar phases. Hydroxylated surface tests show consistent O--H bond-distance distributions across models and minerals, suggesting that these foundation models may provide useful starting points for screening surface hydroxyl stability and volatile-related processes. Performance benchmarks on a single NVIDIA RTX 4090 show that SevenNet-0, MatterSim, and UPET provide the highest throughput among the six tested models, MACE-MH remains practical at intermediate cost, and UMA and NequIP-OAM-L extend the comparison to newer foundation potentials at higher runtime cost and memory demand. These results provide an initial benchmark for applying universal foundation models to lunar mineral simulations and identify key directions for future ab initio validation, model fine-tuning, and applications to lunar volatile evolution, space weathering, ISRU, and polar sample return studies.
 
\end{abstract}

\section{Introduction}

The Moon is entering a new era of exploration and sample-return science, driven by missions such as NASA's Artemis program, China's Chang'e series, and various commercial landers. These efforts renew the need to characterize lunar regolith both as a record of the lunar space environment and as a primary feedstock for \textit{in-situ} resource utilization (ISRU)~\cite{zhang2023overview}, especially with samples returned by the Apollo and Chang'e missions.
Molecular dynamics (MD) simulations have emerged as a powerful approach for investigating atomic-scale processes in returned samples, including the effects of billions of years of micrometeoroid bombardment, solar-wind irradiation, and surface charging. These processes include solar wind implantation, dielectric breakdown, impact-induced redox reactions, and nanophase iron formation~\cite{huang2021molecularMM,huang2022molecular,huang2021molecularDB,huang2025micrometeoroid,huang2026coupled,huang2026revealing,shoji2025reactive,shoji2026molecular,georgiou2025effect}. However, previous reactive MD studies have been largely limited to simplified systems, such as SiO$_2$-rich materials or Fe$_2$SiO$_4$, primarily because traditional reactive methods such as ReaxFF \cite{van2001reaxff} depend on the availability and transferability of suitable force-field parameter sets, both of which remain limited.

The limitations of current computational methods are particularly evident when considering the mineralogical heterogeneity revealed by recent analyses of Apollo and Chang'e samples. These studies highlight that distinct mineral species exhibit unique properties relevant to volatile trapping and redox chemistry. Representative phases include \textit{forsterite} (Mg$_2$SiO$_4$), the Mg-end member of olivine; \textit{fayalite} (Fe$_2$SiO$_4$), the Fe-end member and a key redox-sensitive phase in mare materials; \textit{ilmenite} (FeTiO$_3$), critical for reduction chemistry and oxygen production; and \textit{anorthite} (CaAl$_2$Si$_2$O$_8$), the framework silicate dominant in highland terrains~\cite{smyth1975fayalite,kiefer2012lunar,wechsler1984ilmenite,wainwright1971anorthite}. Modeling these minerals across diverse silicate, oxide, and hydrogen-bearing environments requires a universal force field capable of maintaining accuracy without the need for material-specific reparametrization.

To address this requirement for chemical universality, machine-learning interatomic potentials (MLIPs) trained on density functional theory (DFT) data offer a promising solution. These models provide the computational efficiency of classical MD while retaining the many-body descriptions of local atomic environments characteristic of \textit{ab initio} methods~\cite{deringer2021machine,batzner2022nequip}. Lately, universal or foundation MLIPs have been trained on vast datasets spanning much of the periodic table, theoretically enabling multi-element simulations without further development. However, these models are generally optimized for board materials science and chemistry applications. Their reliability in describing lunar minerals—characterized by space weathering processes, and specific Fe- and Ti-bearing redox states—remains unverified. Consequently, domain-specific benchmarking is necessary to evaluate their transferability and identify failure modes before they can be reliably applied to lunar science problems.

In this work, we address this validation gap by benchmarking six universal foundation models, MACE-MH~\cite{batatia2024foundation}, MatterSim~\cite{yang2024mattersim}, SevenNet-0~\cite{park2024sevennet}, UPET~\cite{mazitov2025pet}, UMA~\cite{wood2026family}, and NequIP-OAM-L~\cite{tan2026high,kavanagh_2025_17087883}, using NVT molecular dynamics simulations of forsterite, fayalite, ilmenite, and anorthite. We assess structural fidelity through bond distances, bond angles, and partial radial distribution functions against crystallographic references, while also profiling GPU wall-clock performance and memory utilization on an NVIDIA RTX 4090. Additionally, we evaluate hydroxylated surfaces of these four minerals to test predicted -OH bond distances, providing an initial assessment of model applicability to volatile-related processes. By validating these foundation models for lunar mineralogy, this work establishes a rigorous baseline for expanding atomistic modeling into large-scale simulations of regolith evolution and comparative studies of diverse landing sites. Ultimately, these benchmarks offer the foundational validation needed to develop a new class of high-fidelity simulations that can bridge the gap between microscopic chemical reactions and macroscopic planetary processes, supporting the design of ISRU systems and the interpretation of future polar sample return missions.
%%============================================================
\section{Methods}
\label{sec:methods}
%%============================================================

\subsection{Mineral structures and supercells}
\label{ssec:structures}

\begin{figure}[htbp]
  \centering
  \includegraphics[width=0.8\textwidth]{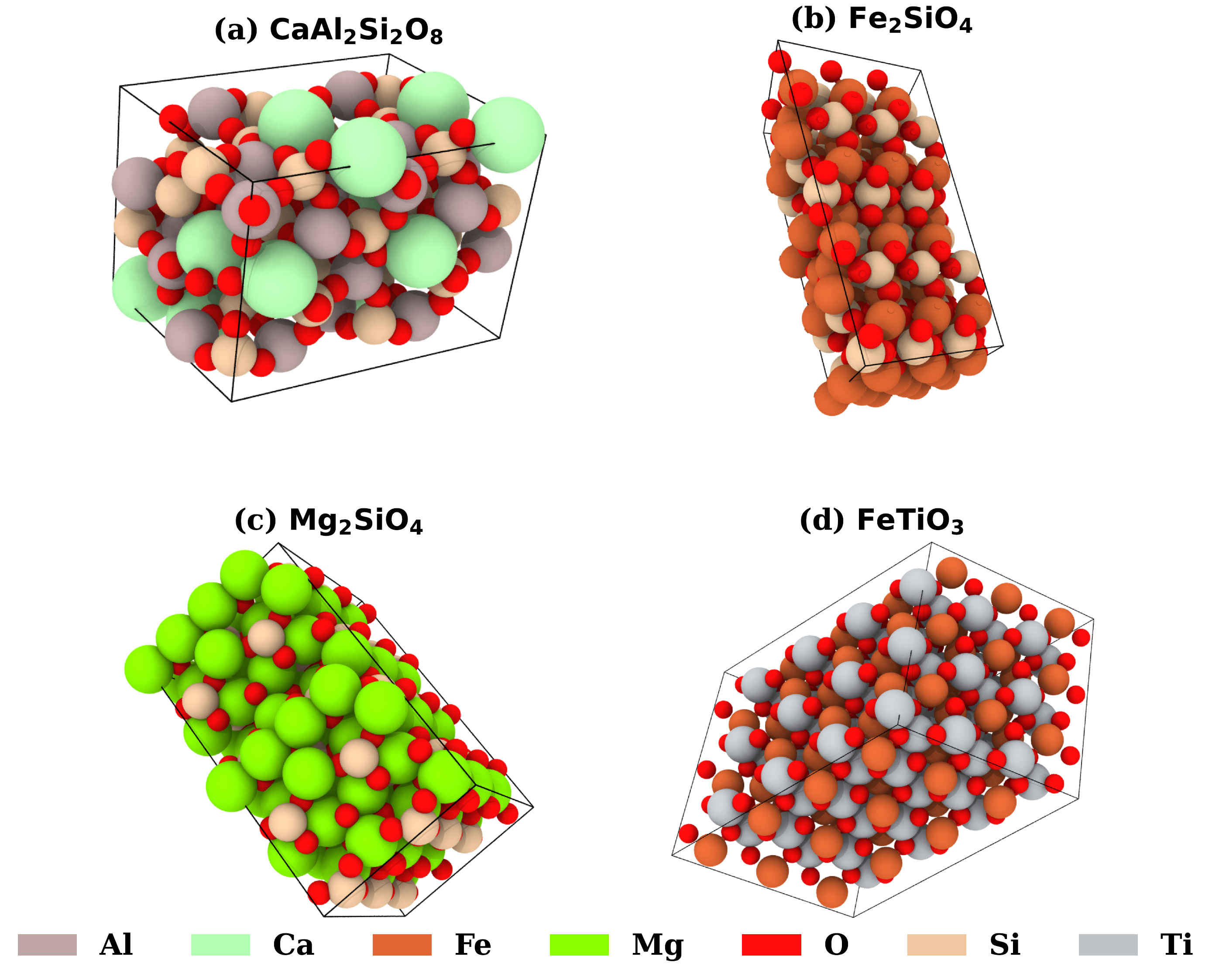}
  \caption{Crystal structures of the four lunar regolith minerals studied.
           Atomic positions are shown as spheres coloured by element
           (Mg: green; Fe: brown; Ti: purple; Ca: Cyan; Al: orange;
           Si: gold; O: red).
           From left to right: anorthite (CaAl$_2$Si$_2$O$_8$, 208 atoms), forsterite (Mg$_2$SiO$_4$, 336 atoms),
           fayalite (Fe$_2$SiO$_4$, 336 atoms),
           ilmenite (TiFeO$_3$, 360 atoms).
           Supercell boundaries are indicated by black lines.}
  \label{fig:structures}
\end{figure}

Crystal structures for all four minerals were retrieved from the Materials
Project~\cite{jain2013commentary} and compared with published experimental
diffractometry data before supercell construction.
Supercells were constructed using the Atomic Simulation Environment
(ASE)~\cite{larsen2017atomic} and inspected in VESTA.
For \textit{forsterite} (Mg$_2$SiO$_4$, space group $Pbnm$, $Z=4$), a
$3\times 2\times 2$ supercell of the conventional orthorhombic cell yielded
336 atoms (84 formula units), with equilibrium lattice parameters
($a = 4.762$\,\AA, $b = 10.225$\,\AA, $c = 5.994$\,\AA) matching the
single-crystal diffraction study of Hazen~\cite{hazen1976forsterite}.
An identical $3\times 2\times 2$ supercell was used for isostructural
\textit{fayalite} (Fe$_2$SiO$_4$, space group $Pbnm$, $Z=4$), using
parameters from Smyth~\cite{smyth1975fayalite}
($a = 4.820$\,\AA, $b = 10.477$\,\AA, $c = 6.105$\,\AA).
For \textit{ilmenite} (FeTiO$_3$, space group $R\bar{3}$, $Z=6$), we
constructed a $3\times 3\times 2$ hexagonal supercell containing 360 atoms
(60 formula units) based on Wechsler and Prewitt~\cite{wechsler1984ilmenite}
($a = 5.088$\,\AA, $c = 14.093$\,\AA).
For \textit{anorthite} (CaAl$_2$Si$_2$O$_8$, space group $P\bar{1}$, $Z=8$),
a $2\times 2\times 2$ triclinic supercell containing 208 atoms (16 formula
units) was derived from Wainwright and Starkey~\cite{wainwright1971anorthite}.
Figure~\ref{fig:structures} illustrates the resulting simulation cells.

\subsection{Foundation models}
\label{ssec:models}

To assess the impact of structural priors and training scale on predictive accuracy, we evaluated six universal machine-learning interatomic potentials encompassing equivariant message-passing, invariant graph neural networks, and transformer architectures. For the MACE family, we used the most up-to-date MACE-MH model. The MACE architecture employs an $\mathrm{E}(3)$-equivariant message-passing network utilizing Clebsch--Gordan tensor products to construct strictly equivariant many-body features. Similarly relying on geometric equivariance, SevenNet-0 (7net-omni-v1)~\cite{park2024sevennet} utilizes the $\mathrm{SE}(3)$-equivariant SO3Net architecture. Comprising 843,000 parameters across five convolutional layers with a 5\,\AA\ cutoff, this model was also trained on the MPtrj dataset but operates efficiently in \texttt{float32} precision.

In addition to strictly equivariant networks, we also tested models leveraging invariant graphs and attention mechanisms. MatterSim v1.0~\cite{yang2024mattersim} implements the M3GNet architecture~\cite{chen2022m3gnet}, an invariant graph neural network that captures three-body interactions through explicit bond-angle terms. Operating in \texttt{float32} precision, its 890,000 parameters are distributed across three interaction blocks. MatterSim benefits from massive-scale pretraining on over 17 million DFT calculations encompassing both the Materials Project and the OMAT24 datasets, granting it broad coverage across complex phase spaces. UPET (PET-OMAT-M v1.0)~\cite{mazitov2025pet} represents a Point Edge Transformer approach wherein each interaction block functions as a full multi-head self-attention layer treating atomic environments as token sequences. We also included UMA, a newer family of Universal Models for Atoms trained across molecular, materials, and catalytic domains, and NequIP-OAM-L, an OMat/Alexandria-trained NequIP foundation potential for inorganic materials. 

\subsection{Molecular dynamics protocol}
\label{ssec:md}

All dynamic simulations were carried out within the ASE framework~\cite{larsen2017atomic}, with each foundation model used as the energy and force calculator. For every mineral--model combination, the simulation proceeded in two phases. First, the constructed supercell geometry was relaxed using the FIRE algorithm~\cite{guenole2020assessment} until the maximum force was below $0.05 \mathrm{eV},\text{\AA}^{-1}$. Second, the relaxed structure was propagated in the NVT ensemble using a Langevin thermostat at $T = 300$\,K. The equations of motion were integrated using a timestep of $\Delta t = 1.0$ fs for 1000 steps, corresponding to a total simulation time of 1 ps. This NVT run was also used as a short-timescale stability benchmark to ensure that no obvious dynamical instabilities occurred during the simulation. Specifically, we checked whether each model--mineral combination could maintain a physically reasonable structure without atom ejection, unphysical bond-length divergence, or rapid structural collapse. The cell vectors were held fixed to the crystallographic reference volumes, so the NVT setup provides a local structural and short-timescale dynamical benchmark rather than an equation of state or thermal-expansion calculation. Periodic boundary conditions and the minimum-image convention were used throughout.

\subsection{Structural analysis}
\label{ssec:analysis}

Structural fidelity was evaluated from first-shell coordination geometries and
medium-range periodic order.
First-shell bond distances were extracted across the last 500 stable MD frames using fixed
species-pair cutoffs: Si--O 2.0\,\AA, Mg--O 2.8\,\AA, Fe--O 2.8\,\AA, Ti--O
2.6\,\AA, Al--O 2.4\,\AA, and Ca--O 3.2\,\AA.
For each frame, the minimum-image distance was calculated for every pair within
the relevant cutoff, and the reported mean and standard deviation were computed
over the trajectory.
Bond angles were computed from the final 500 simulation frames by identifying
first-shell neighbors with the same cutoff definitions and evaluating the angle
subtended at the central cation.
Partial radial distribution functions, $g_{\alpha\beta}(r)$, were compiled
from the final 500 frames of each trajectory, representing the equilibrated
second half of the simulation.
These distributions were binned at 0.02\,\AA\ resolution and normalized to the
ideal-gas limit using the stoichiometric number density for each pair type.
The fixed-cutoff analysis is intended as a systematic structural comparison;
coordination changes near cutoff boundaries should therefore be interpreted
qualitatively, particularly for the broad Fe--O and Ti--O distributions. 

\subsection{Performance Benchmark}

To profile computational efficiency, GPU wall-clock time was measured with
Python's \texttt{time.perf\_counter} around the \texttt{get\_forces()} call at
each integration step, after model loading and structure initialization.
Peak GPU memory allocation was tracked using
\texttt{torch.cuda.max\_memory\_allocated()} where supported by the model
backend.
All calculations were executed on an NVIDIA GeForce RTX 4090
(24\,GB GDDR6X) using CUDA 12.4 and PyTorch 2.3.
Because not all model wrappers expose identical memory instrumentation, missing
peak-memory entries are reported as unavailable rather than inferred.

%%============================================================
\section{Results and Discussions}
\label{sec:results}
%%============================================================

\subsection{NVT molecular dynamics stability check}
\label{ssec:stability}

\begin{figure}[htbp]
  \centering
  \includegraphics[width=\textwidth]{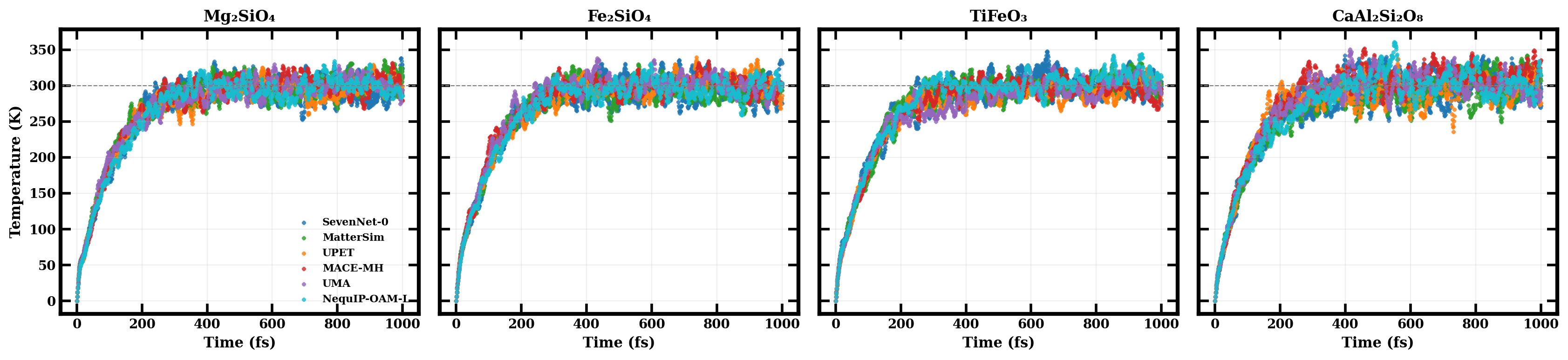}
  \caption{Temperature (top row) and per-atom potential energy (bottom row)
           as a function of simulation time for all six foundation models
           on each mineral.
           The dashed grey line marks the target temperature of 300\,K.
           All models reach thermal equilibrium within 100--200\,fs and
           maintain stable dynamics throughout the 1\,ps trajectory.
           Model colours: SevenNet-0 (blue), MatterSim (green),
           UPET (orange), MACE-MH (red), UMA (purple), and NequIP-OAM-L (cyan).}
  \label{fig:temperature}
\end{figure}

All 24 model--mineral combinations (six models $\times$ four minerals) completed 1000 NVT MD steps without structural instabilities, such as atom ejection or bond-length divergence. The successful completion of these trajectories serves as a critical validation of the models' foundational stability in lunar regolith modeling, demonstrating their immediate transferability to these complex mineral compositions without requiring specialized retraining. Temperature traces during NVT-MD (Figure~\ref{fig:temperature}) show that all models equilibrate within approximately 100--200 steps and maintain a stable 300~K mean with fluctuations appropriate to the finite system sizes ($\delta T \approx \pm 20$--30\,K, consistent with the equipartition expectation $\delta T / T = (2/3N)^{1/2}$).

%%%%%%%%%%%%%%%%%%%%%%%%%%%%%%%%%%%%%%%%%%%%%%%%
\subsection{Radial distribution functions}
\label{ssec:rdf}

\begin{figure}[htbp]
\centering
\includegraphics[width=\textwidth]{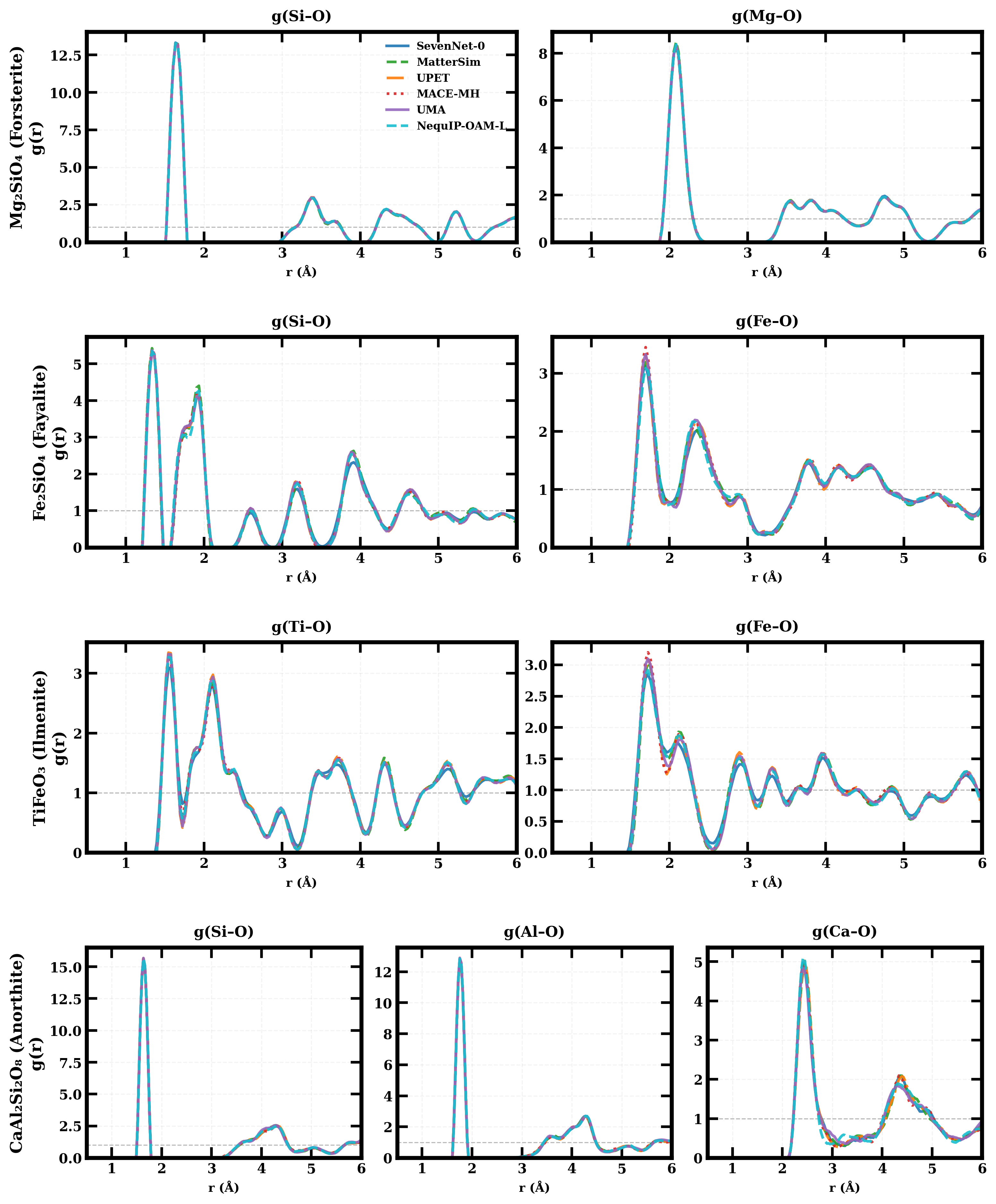}
\caption{Partial radial distribution functions $g_{\alpha\beta}(r)$
computed from the last 500 frames of each trajectory using a bin
width of 0.03,\AA.
Each panel corresponds to one mineral, and columns within each
panel show chemically distinct pair types.
The dashed horizontal line at $g(r)=1$ marks the ideal-gas
uncorrelated limit.
Sharp first-shell peaks are observed for Si--O and Al--O pairs,
confirming the integrity of tetrahedral network bonding.
Broader first-shell peaks for Fe--O and Ti--O pairs indicate
larger short-timescale fluctuations in octahedral coordination
environments.
Top to bottom: forsterite (Mg$_2$SiO$_4$), fayalite
(Fe$_2$SiO$_4$), ilmenite (TiFeO$_3$), and anorthite
(CaAl$_2$Si$_2$O$_8$).}
\label{fig:gr}
\end{figure}

Partial radial distribution functions $g_{\alpha\beta}(r)$ were computed from the second half of each NVT trajectory to assess whether the six foundation models preserve consistent short-range order during the 300\,K simulations. As shown in Figure~\ref{fig:gr}, the Mg-bearing and framework silicates exhibit sharp and well-defined first coordination shells. For forsterite, the Si--O $g(r)$ shows a narrow and symmetric first peak at $r \approx 1.65$\,\AA\  across the six-model set, with negligible differences in peak position, height, or width. The second coordination shell near $r \approx 2.5$\,\AA\ is also reproduced consistently, indicating that the tetrahedral silicate framework remains stable over the short NVT trajectory. Similarly, for anorthite, the Si--O and Al--O first-shell peaks are sharp and well resolved. The slight shoulder near $r \approx 1.80$\,\AA\ reflects the longer Al--O bonds in the aluminosilicate framework and is consistent with the bond-distance analysis in Section~\ref{ssec:bonds_minerals}.

In contrast, the Fe- and Ti-bearing minerals show broader first-shell distributions. In fayalite, the Fe--O $g(r)$ has a noticeably broader first peak centered near $r \approx 2.18$\,\AA, with larger intermodel differences in peak height than those observed for forsterite and anorthite. In ilmenite, the Ti--O and Fe--O peaks are also substantially broader than the Si--O and Al--O peaks in the silicate frameworks. However, the first-shell peaks remain distinct and well separated from the next coordination shells, indicating that no gross structural rearrangement occurs during the short MD simulations. Overall, the RDFs show that the six-model set gives broadly consistent short-range structure for the ordered mineral lattices, while Fe--O and Ti--O coordination environments exhibit larger dynamical broadening and therefore require more careful validation.

\subsection{Bond-distance benchmarks across lunar minerals}
\label{ssec:bonds_minerals}

\begin{figure}[htbp]
\centering
\includegraphics[width=\textwidth]{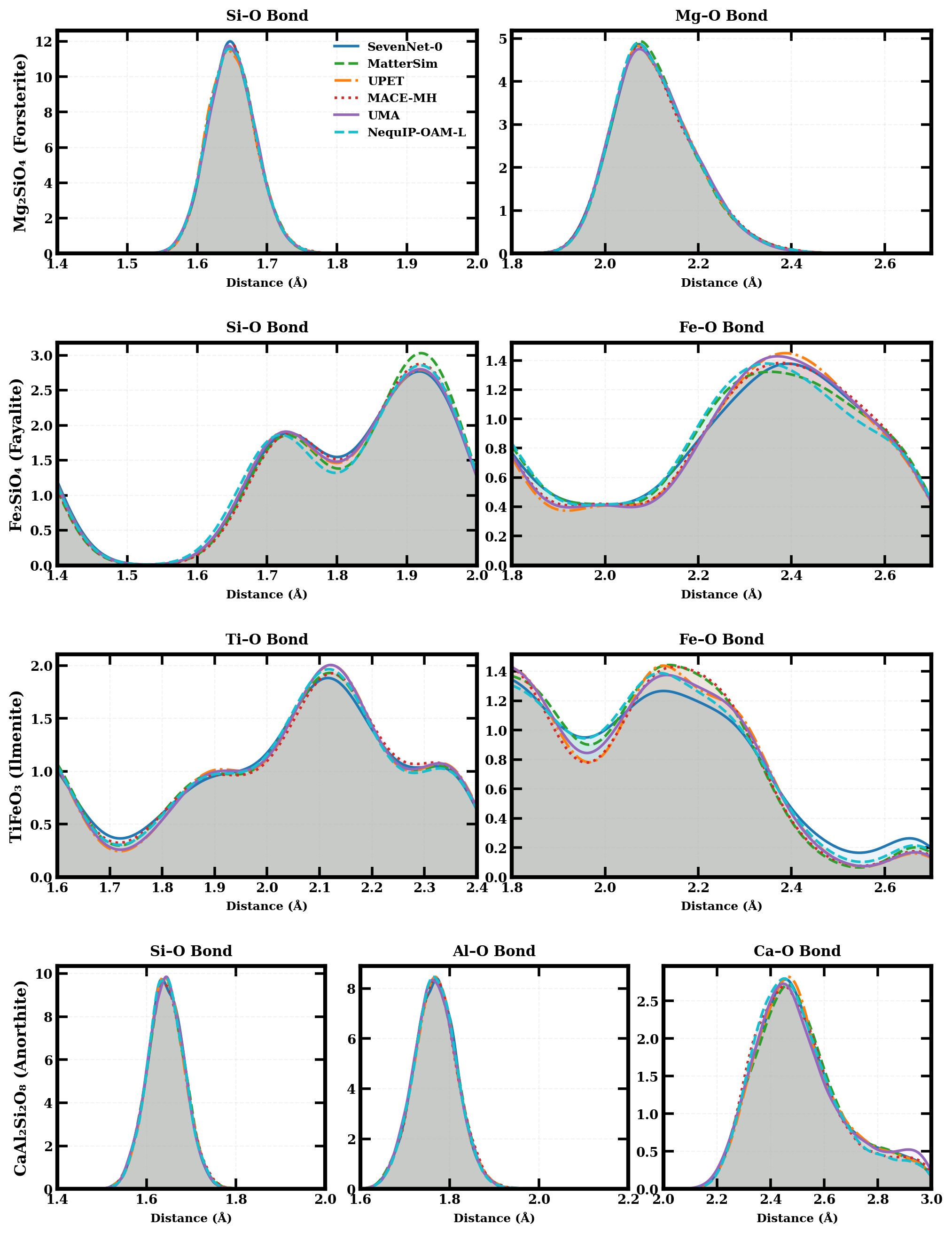}
\caption{Bond-distance probability distributions from the final 500 frames for all six models and minerals.
Distances are sampled every 5 frames using species-pair cutoffs
to select first-shell pairs only: Si--O, 2.0,\AA; Mg/Fe--O,
2.7,\AA; Ti--O, 2.4,\AA; Al--O, 2.2,\AA; and Ca--O,
3.0,\AA.
The Si--O and Al--O distributions are tightly peaked, while the
Fe--O and Ti--O distributions are substantially broader, consistent
with the RDF analysis in Figure~\ref{fig:gr}}
\label{fig:bonds}
\end{figure}

The bond-distance statistics provide a quantitative comparison of the first-shell coordination environments identified from the RDFs. Table~\ref{tab:bonds} summarizes the mean nearest-neighbor bond distances and standard deviations for all six models from the 300~K MD trajectories, while Figure~\ref{fig:bonds} shows the corresponding probability distributions for each model. The reference values in Table~\ref{tab:bonds} are static first-shell ranges recomputed from the corresponding crystallographic starting supercells described in section 2.1. They therefore describe site inequivalence in the static structures rather than finite-temperature AIMD or experimental thermal broadening.

Across the Mg-bearing silicate and aluminosilicate structures, the intermodel differences are small and the MD means remain close to the static first-shell ranges. In forsterite, all six models predict $Si\text{--}O$ distances of $1.650$\,\AA\ with $\sigma \approx 0.034$\,\AA, within the static reference range of $1.617\text{--}1.658$\,\AA, and $Mg\text{--}O$ distances of approximately $2.109$\,\AA, within the corresponding $2.041\text{--}2.207$\,\AA\ range. In anorthite, the models also preserve the distinction between tetrahedral $Si\text{--}O$ and $Al\text{--}O$ bonding, with six-model mean ranges of $1.643\text{--}1.644$\,\AA\ for $Si\text{--}O$ and $1.767\text{--}1.769$\,\AA\ for $Al\text{--}O$. The broader $Ca\text{--}O$ distribution is also consistent with the wide static range of $2.324\text{--}2.962$\,\AA\ expected for distorted multi-fold Ca coordination.

The largest deviations occur in the Fe-bearing minerals. In fayalite, the model set predicts $Si\text{--}O$ distances of $1.714\text{--}1.718$\,\AA, which are longer than the static reference range of $1.629\text{--}1.653$\,\AA. The $Fe\text{--}O$ mean distances ($2.178\text{--}2.182$\,\AA) remain inside the static first-shell range of $2.034\text{--}2.303$\,\AA, but the associated standard deviations are much larger than those for $Si\text{--}O$ in forsterite. In ilmenite, $Ti\text{--}O$ distances of $1.984\text{--}1.990$\,\AA\ fall within the static range of $1.863\text{--}2.148$\,\AA, whereas the $Fe\text{--}O$ distances of $1.997\text{--}2.011$\,\AA\ are slightly below the lower edge of the static $2.045\text{--}2.282$\,\AA\ range. These trends are consistent with the RDF analysis, where $Fe\text{--}O$ and $Ti\text{--}O$ first-shell peaks are broader than the $Si\text{--}O$ and $Al\text{--}O$ peaks.

The disproportionately large standard deviations observed in fayalite and ilmenite—compared to the narrower, more rigid distributions in the Mg- and Ca-bearing frameworks—suggest that these structural fluctuations are not solely attributable to thermal broadening at 300 K. Instead, they likely reflect an artificial softening of the potential energy surface around the transition metals \cite{deng2025systematic}. Universal foundation models trained on large, aggregated datasets frequently struggle to consistently capture the strong electron correlation and specific magnetic orderings associated with the localized $d$-electrons of Fe$^{2+}$ and Ti$^{4+}$. Because these models generally lack explicit spin-dependent features, the resulting global fit yields shallower local energy minima than would be expected from targeted, system-specific DFT+U calculations, manifesting as enhanced bond-distance variability during finite-temperature dynamics.

The standard deviations further highlight the difference between framework silicates and Fe-bearing phases. In anorthite, the distributions remain relatively narrow, with $\sigma \approx 0.040\text{--}0.041$\,\AA\ for Si--O, $\sigma \approx 0.047\text{--}0.048$\,\AA\ for Al--O, and $\sigma \approx 0.167\text{--}0.180$\,\AA\ for Ca--O. In contrast, fayalite shows larger fluctuations, with $\sigma \approx 0.235\text{--}0.238$\,\AA\ for Si--O and $\sigma \approx 0.334\text{--}0.340$\,\AA\ for Fe--O. Ilmenite also shows elevated fluctuations for both Ti--O and Fe--O pairs, with $\sigma \approx 0.25\text{--}0.27$\,\AA. Overall, the RDF and bond-distance benchmarks indicate that while these universal models reproduce local bonding in ordered silicate and aluminosilicate frameworks reasonably well, achieving comparable accuracy for Fe-bearing minerals will likely benefit from targeted fine-tuning using additional lunar-relevant iron-bearing ground truth data before deploying them in high-fidelity space-weathering or volatile-evolution simulations.

\begin{table}[htbp]
\caption{Bond distances (mean $\pm$ standard deviation, \AA) from 1000-step
NVT-MD at 300\,K for all four minerals and six models.
Static reference ranges were recomputed from the first-shell distances of the
corresponding starting supercells;
these ranges describe crystallographic/site inequivalence in static structures,
not finite-temperature uncertainties.
Crystallographic sources:
$^a$Hazen (1976)~\cite{hazen1976forsterite};
$^b$Smyth (1975)~\cite{smyth1975fayalite};
$^c$Wechsler and Prewitt (1984)~\cite{wechsler1984ilmenite};
$^d$Wainwright and Starkey (1971)~\cite{wainwright1971anorthite}.}
\label{tab:bonds}
\centering
\scriptsize
\begingroup\color{black}
\setlength{\tabcolsep}{2.2pt}
\begin{tabular}{@{}llcccccc@{}}
\toprule
Mineral & Bond pair
  & SevenNet-0 & MatterSim & UPET & MACE-MH & UMA & NequIP-OAM-L \\
\midrule
\multirow{4}{*}{\shortstack[l]{\textit{Forsterite}\\Mg$_2$SiO$_4$\\(336 atoms)}}
  & Si--O & $1.650\pm0.034$ & $1.650\pm0.034$ & $1.650\pm0.034$ & $1.650\pm0.034$ & $1.650\pm0.034$ & $1.650\pm0.034$ \\
  &       & \multicolumn{6}{c}{\textit{Static ref.$^a$: $1.617$--$1.658$\,\AA}} \\[2pt]
  & Mg--O & $2.109\pm0.088$ & $2.109\pm0.087$ & $2.109\pm0.088$ & $2.109\pm0.089$ & $2.109\pm0.087$ & $2.109\pm0.088$ \\
  &       & \multicolumn{6}{c}{\textit{Static ref.$^a$: $2.041$--$2.207$\,\AA}} \\
\midrule
\multirow{4}{*}{\shortstack[l]{\textit{Fayalite}\\Fe$_2$SiO$_4$\\(336 atoms)}}
  & Si--O & $1.714\pm0.235$ & $1.718\pm0.238$ & $1.714\pm0.236$ & $1.717\pm0.237$ & $1.714\pm0.235$ & $1.714\pm0.237$ \\
  &       & \multicolumn{6}{c}{\textit{Static ref.$^b$: $1.629$--$1.653$\,\AA}} \\[2pt]
  & Fe--O & $2.178\pm0.340$ & $2.181\pm0.336$ & $2.181\pm0.340$ & $2.182\pm0.339$ & $2.182\pm0.340$ & $2.178\pm0.334$ \\
  &       & \multicolumn{6}{c}{\textit{Static ref.$^b$: $2.034$--$2.303$\,\AA}} \\
\midrule
\multirow{4}{*}{\shortstack[l]{\textit{Ilmenite}\\TiFeO$_3$\\(360 atoms)}}
  & Ti--O & $1.984\pm0.268$ & $1.986\pm0.267$ & $1.990\pm0.268$ & $1.988\pm0.268$ & $1.990\pm0.267$ & $1.985\pm0.267$ \\
  &       & \multicolumn{6}{c}{\textit{Static ref.$^c$: $1.863$--$2.148$\,\AA}} \\[2pt]
  & Fe--O & $2.011\pm0.273$ & $2.004\pm0.254$ & $1.998\pm0.258$ & $1.997\pm0.258$ & $1.998\pm0.257$ & $2.007\pm0.262$ \\
  &       & \multicolumn{6}{c}{\textit{Static ref.$^c$: $2.045$--$2.282$\,\AA}} \\
\midrule
\multirow{6}{*}{\shortstack[l]{\textit{Anorthite}\\CaAl$_2$Si$_2$O$_8$\\(208 atoms)}}
  & Si--O & $1.643\pm0.040$ & $1.643\pm0.041$ & $1.643\pm0.041$ & $1.644\pm0.041$ & $1.643\pm0.041$ & $1.643\pm0.040$ \\
  &       & \multicolumn{6}{c}{\textit{Static ref.$^d$: $1.596$--$1.673$\,\AA}} \\[2pt]
  & Al--O & $1.768\pm0.047$ & $1.768\pm0.048$ & $1.768\pm0.048$ & $1.769\pm0.048$ & $1.767\pm0.048$ & $1.768\pm0.047$ \\
  &       & \multicolumn{6}{c}{\textit{Static ref.$^d$: $1.712$--$1.794$\,\AA}} \\[2pt]
  & Ca--O & $2.516\pm0.171$ & $2.518\pm0.171$ & $2.517\pm0.170$ & $2.510\pm0.172$ & $2.519\pm0.180$ & $2.508\pm0.167$ \\
  &       & \multicolumn{6}{c}{\textit{Static ref.$^d$: $2.324$--$2.962$\,\AA}} \\
\bottomrule
\end{tabular}
\endgroup
\end{table}

%%%%%%%%%%%%%%%%%%%%%%%%%%%%%%%%%%%%%%%%%%%%%%%%

\subsection{Bond angle distributions}
\label{ssec:angles}

\begin{figure}[htbp]
  \centering
  \includegraphics[width=\textwidth]{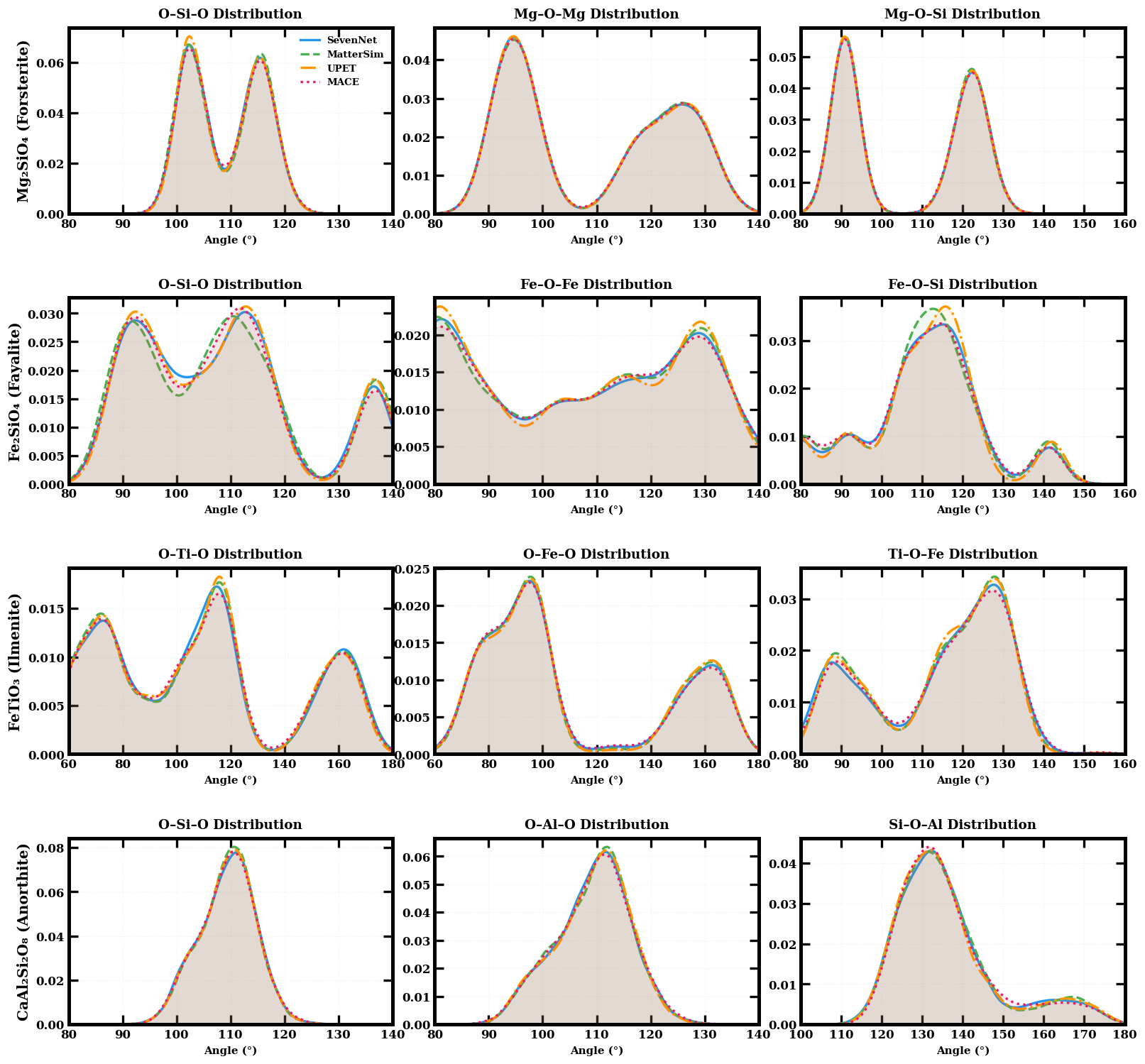}
  \caption{Bond angle probability distributions from the final 500 frames.
           Columns within each panel correspond to chemically
           distinct angle types within each mineral.
           All six models reproduce O--Si--O and O--Al--O tetrahedral geometry
           within $\pm 0.8^\circ$ of each other.
           The broader O--Ti--O and O--Fe--O distributions in ilmenite
           reflect the distorted octahedral environment of the
           corundum-type structure.
           Top to bottom: forsterite, fayalite, ilmenite, and anorthite.}
  \label{fig:angles}
\end{figure}

Bond angle distributions (Figure~\ref{fig:angles}) provide complementary
information about local coordination geometry.
For the critical O--Si--O tetrahedral angle, all six models predict values
very close to the ideal tetrahedral angle of 109.47$^\circ$:
in forsterite the range across models is 108.7$^\circ$--109.4$^\circ$,
and in anorthite it is 109.0$^\circ$--109.3$^\circ$.
The Mg--O--Mg bridge angle in forsterite is 109.2$^\circ$--109.4$^\circ$,
consistent with the near-regular octahedral arrangement of oxygen atoms
around Mg in the M1 site.
The O--Ti--O angle in ilmenite ($103.9^\circ$--$104.0^\circ$) reflects the
face-sharing octahedral geometry of the corundum-derived structure,
which deviates from the 90$^\circ$ ideal of a perfect octahedron.
The Si--O--Al bridging angle in anorthite ($135.6^\circ$--$136.4^\circ$)
describes the flexibility of the Si/Al tetrahedral framework and is in good
agreement with crystallographic values in the range $130^\circ$--$140^\circ$
for anorthitic plagioclase.
The O--Al--O intratetrahedral angle ($109.0^\circ$--$109.3^\circ$) confirms
near-ideal aluminium tetrahedral geometry in all models.

\subsection{Surface hydroxyl bond distances}
\label{ssec:oh_distance}

To evaluate the capability of these foundation models to describe hydrogen-bearing surface species, we tested hydroxylated surfaces of the four representative lunar minerals. Hydroxylated models were generated by cleaving one-layer slabs from the benchmark mineral supercells with 30 {\AA} of vacuum: (010) surfaces for forsterite, fayalite, and anorthite, and a (001) surface for ilmenite. Non-bridging surface oxygen atoms—defined as those within 1.0 {\AA} of the highest O position and having no more than one Si/Al neighbor (using Si--O and Al--O cutoffs of 2.0 and 2.2,{\AA}, respectively)—were protonated by placing a H atom along the outward surface normal with an initial O--H distance of 0.98,{\AA}.

\begin{figure}[htbp]
\centering
\includegraphics[width=\textwidth]{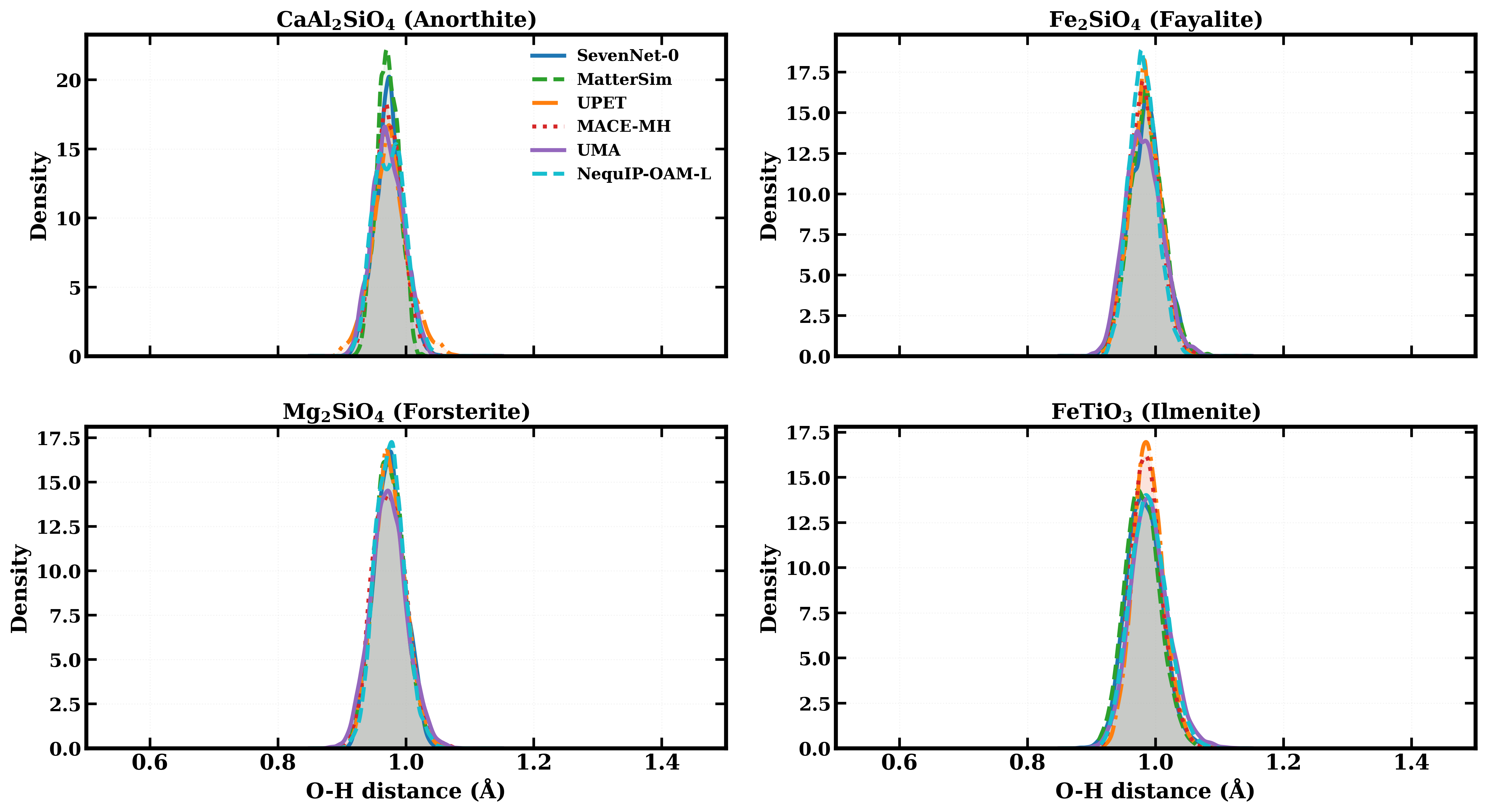}
\caption{O--H bond-distance distributions for hydroxylated surfaces of representative lunar minerals predicted by the six foundation models.
The distributions show strong consistency across models and mineral surfaces, indicating stable surface hydroxyl groups during the short NVT trajectories.
Top to bottom: forsterite (Mg$_2$SiO$_4$), fayalite
(Fe$_2$SiO$_4$), ilmenite (TiFeO$_3$), and anorthite
(CaAl$_2$Si$_2$O$_8$).}
\label{fig:oh_distance}
\end{figure}

Figure~\ref{fig:oh_distance} illustrates the resulting O--H bond-distance distributions. In contrast to the model-dependent fluctuations observed in certain Fe--O and Ti--O coordination environments, the O--H distributions remain remarkably consistent across all six foundation models and mineral surfaces. The analysis used the same stable hydroxylated surface structures and tracked O--H pairs over the final 500 trajectory frames. The preservation of narrow bond-distance distributions during NVT trajectories indicates that these models can robustly represent stable hydroxyl groups without experiencing bond elongation, dissociation, or unphysical motion. This high degree of consistency is a critical prerequisite for enabling large-scale computational studies of lunar volatile evolution, providing a reliable framework for the high-throughput screening of hydrogen retention, migration, and release across diverse mineralogical contexts.

By demonstrating that universal foundation models can reproduce stable O--H bonding environments, this work establishes their potential for accelerating the exploration of mineral-dependent volatile cycling and surface hydroxyl stability. While further \textit{ab initio} validation remains necessary for modeling complex reaction pathways—such as proton transfer, hydroxyl recombination, or water formation—these results suggest that these models are already well-suited for characterizing the initial states of hydroxylated lunar surfaces. This capability significantly expands the scope of possible atomistic investigations into how solar-wind-implanted hydrogen interacts with the Moon's heterogeneous surface, bridging the gap between fundamental mineralogy and macroscopic volatile transport models.

\subsection{Computational performance}
\label{ssec:performance}

\begin{figure}[htbp]
  \centering
  \includegraphics[width=\textwidth]{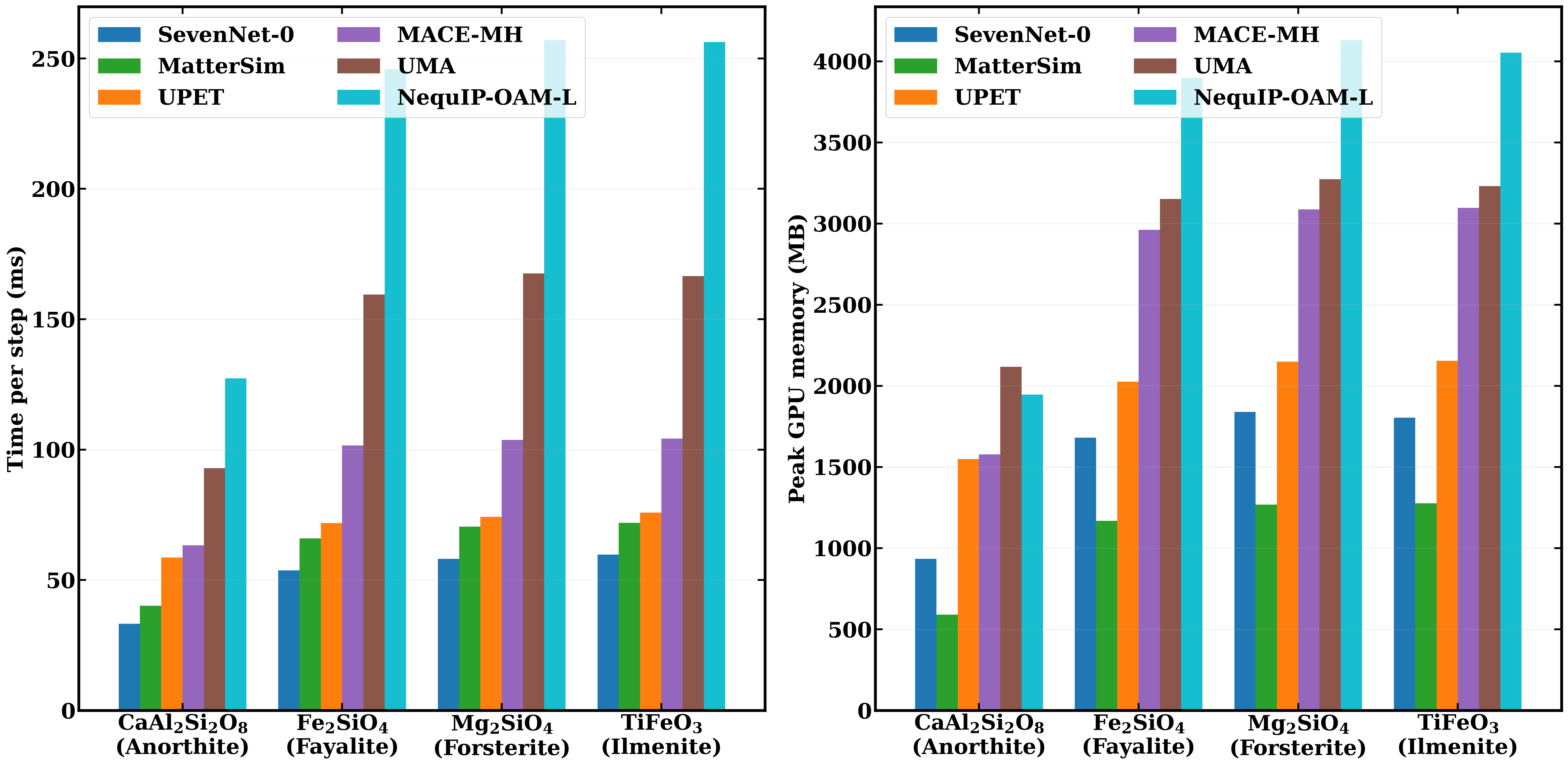}
  \caption{Wall-clock timing comparison on an NVIDIA GeForce RTX 4090 (24\,GB).
           (a)~Total wall-clock time (s) for 1000 NVT-MD steps per mineral
               and model, shown as grouped bars.
           (b)~Mean time per MD step (ms/step) averaged across all four
               mineral supercells.
            SevenNet-0 and MatterSim achieve the highest throughput,
           followed by UPET and MACE-MH, while UMA and NequIP-OAM-L provide
           broader new-model coverage at higher wall-clock cost.} 
  \label{fig:performance}
\end{figure}

We benchmarked the computational performance of the six foundation models by measuring the wall-clock time for 1000 NVT-MD steps and the peak GPU memory allocation for each model--mineral combination. All calculations were carried out on a single NVIDIA GeForce RTX 4090 GPU with 24 GB of memory. Table~\ref{tab:performance} summarizes the timing and memory results, and Figure~\ref{fig:performance} compares the total simulation time and average time per MD step across the four mineral supercells.

\begin{table}[htbp]
\caption{Wall-clock time (seconds for 1000 MD steps) and peak GPU memory (MB) for all six models on each mineral supercell. All calculations were performed on a single NVIDIA GeForce RTX 4090 (24\,GB).}
\label{tab:performance}
\centering
\begingroup\color{black}
\begin{tabular}{llrr}
\toprule
Mineral (atoms) & Model & Time (s) / 1000 steps & Peak GPU (MB) \\
\midrule
\multirow{6}{*}{Mg$_2$SiO$_4$ (336)} & SevenNet-0 & 58.1 & 1839 \\
 & MatterSim & 70.5 & 1268 \\
 & UPET & 74.2 & 2149 \\
 & MACE-MH & 103.7 & 3088 \\
 & UMA & 167.5 & 3274 \\
 & NequIP-OAM-L & 257.0 & 4130 \\
\midrule
\multirow{6}{*}{Fe$_2$SiO$_4$ (336)} & SevenNet-0 & 53.7 & 1680 \\
 & MatterSim & 65.9 & 1168 \\
 & UPET & 71.8 & 2025 \\
 & MACE-MH & 101.6 & 2961 \\
 & UMA & 159.5 & 3152 \\
 & NequIP-OAM-L & 245.9 & 3896 \\
\midrule
\multirow{6}{*}{TiFeO$_3$ (360)} & SevenNet-0 & 59.8 & 1803 \\
 & MatterSim & 71.9 & 1277 \\
 & UPET & 75.9 & 2154 \\
 & MACE-MH & 104.2 & 3097 \\
 & UMA & 166.6 & 3231 \\
 & NequIP-OAM-L & 256.3 & 4053 \\
\midrule
\multirow{6}{*}{CaAl$_2$Si$_2$O$_8$ (208)} & SevenNet-0 & 33.2 & 934 \\
 & MatterSim & 40.1 & 590 \\
 & UPET & 58.6 & 1549 \\
 & MACE-MH & 63.3 & 1578 \\
 & UMA & 92.9 & 2118 \\
 & NequIP-OAM-L & 127.3 & 1947 \\
\bottomrule
\end{tabular}
\endgroup
\vspace{0.5ex}

\end{table}

SevenNet-0 gives the shortest wall-clock time for all four minerals, ranging from 33.2\,s for the 208-atom anorthite supercell to 59.8\,s for the 360-atom ilmenite supercell. MatterSim is the second fastest model, requiring 40.1--71.9\,s for 1000 MD steps, followed by UPET with 58.6--75.9\,s. MACE-MH requires 63.3--104.2\,s, UMA requires 92.9--167.5\,s, and NequIP-OAM-L requires 127.3--257.0\,s for the same simulations. This difference indicates that model architecture and implementation details can have a large effect on practical MD throughput, even when all models are applied to the same mineral structures on the same GPU.

The peak GPU memory usage remains moderate for all tested systems.  For the largest supercell, TiFeO$_3$ with 360 atoms, the peak memory allocation is 1803\,MB for SevenNet-0, 1277\,MB for MatterSim, 2154\,MB for UPET, 3097\,MB for MACE-MH, 3231\,MB for UMA, and 4053\,MB for NequIP-OAM-L.  
% These values are all well below the 24,GB capacity of the RTX 4090. 
MatterSim has the lowest peak memory demand among the six models, while NequIP-OAM-L requires the largest working memory during MD. Overall, the performance benchmark shows that SevenNet-0, MatterSim, and UPET are computationally efficient for short-timescale MD screening of ordered lunar mineral supercells, while MACE-MH, UMA, and NequIP-OAM-L provide the same type of calculation at progressively higher computational cost.
 Assuming approximately linear memory scaling for fixed-density supercells, the 360-atom benchmark suggests that a 24~GB RTX 4090 could support simulations ranging from roughly $2\times10^3$ atoms for the most memory-demanding model, NequIP-OAM-L, to more than $6\times10^3$ atoms for the most memory-efficient model, MatterSim. These estimates should be interpreted as approximate memory-based upper bounds rather than strict limits, because the actual maximum system size depends on the cutoff radius, neighbor-list size, model architecture, CUDA overhead, and runtime settings.

%%============================================================
\section{Conclusions}
\label{sec:conclusions}
%%============================================================

 This work benchmarks six universal foundation models, MACE-MH, MatterSim, SevenNet-0, UPET, UMA, and NequIP-OAM-L, for short-timescale molecular dynamics simulations of representative lunar minerals.  All model--mineral combinations completed 1000 NVT steps at 300\,K without atom ejection, bond-length divergence, or rapid structural collapse, providing an initial stability check for simulations of lunar-relevant compositions. The RDFs, bond-distance distributions, and bond-angle analysis show that the models give broadly consistent local structures for ordered Mg-bearing silicates and aluminosilicate frameworks. In particular, Si--O and Al--O tetrahedral environments are well preserved, and the predicted bond distances are generally close to crystallographic reference values. The main limitations appear in Fe- and Ti-bearing phases, where Fe--O and Ti--O coordination environments show broader distributions and larger short-timescale fluctuations. This indicates that universal foundation models can be useful for initial screening of ordered lunar minerals, but Fe-bearing and Ti-bearing systems require more validation as well as fine-tuning before being used for detailed studies of redox chemistry, space weathering, or ISRU-relevant reaction pathways.

 The hydroxylated surface tests further show that the predicted O--H bond distances are highly consistent across the six models and four mineral surfaces, suggesting that these models may provide a useful starting point for studying surface hydroxyl stability and hydrogen-bearing species in lunar minerals.  This capability is important for future simulations of hydrogen retention, volatile cycling, and mineral-dependent surface chemistry, especially in the context of Artemis and future polar sample return missions. At the same time, the present benchmark is limited to short NVT trajectories, ordered crystal structures, and simple hydroxylated surfaces. 
Looking forward, extending these benchmarks to a broader range of non-equilibrium conditions would be a valuable step toward fully capturing lunar space-weathering mechanisms. For instance, investigating point defects and amorphous, radiation-damaged structures could offer deeper insights into solar wind and ion irradiation processes. Similarly, evaluating these models under high-pressure regimes would help assess their capability to simulate the extreme conditions of micrometeoroid impacts. Exploring these disordered states, along with complex regolith models and temperature-dependent volatile kinetics, represents an exciting avenue for bridging the gap between universal foundation models and high-fidelity planetary science simulations.
Direct comparison with DFT calculations will be needed for proton transfer, hydroxyl recombination, water formation, desorption, defect trapping, and redox reactions. Fine-tuning foundation models with lunar-relevant ab initio datasets may provide a practical path toward more reliable simulations of volatile evolution, space weathering, and ISRU processes across the mineralogical diversity of lunar regolith.

%%============================================================

\ack{We acknowledge useful feedback and discussions from Brant M. Jones and Shuai Li. This research was supported in part through research cyber infrastructure resources and services provided by the Partnership for an Advanced Computing Environment (PACE) at the Georgia Institute of Technology, Atlanta, Georgia, USA. Z.H. is supported by SSERVI-CLEVER (NNH22ZDA020C/80NSSC23M022), and Astrobiology Fellowship.
}

\section*{Conflict of interest statement}

The authors declare no competing interests.

\roles{Z.H. \& K.N.: Conceptualization, Methodology, Software, Validation, Formal
Analysis, Investigation, Data Curation, Writing -- Original Draft, Writing --
Review \& Editing, Visualization
}

\data{The code and input data supporting this study are publicly available at
\url{https://github.com/ziyuhuang652/Benchmark_FoundationModel_LunarRegolith}.
The repository contains the bulk mineral structures, hydroxylated surface structures, source CIF/reference files, benchmark-running scripts, memory-profiling scripts, O--H analysis scripts, analysis notebooks, environment setup files, model-loading workflows, and a data manifest. Generated trajectories, intermediate output tables, figures, and manuscript build products are not included in the repository but can be regenerated using the provided scripts. The foundation model codes and weights are available from their original public sources: MACE-MH at \url{https://github.com/ACEsuit/mace}; MatterSim at \url{https://github.com/microsoft/mattersim}; SevenNet at \url{https://github.com/MDIL-SNU/SevenNet}; UPET/PET at \url{https://github.com/lab-cosmo/pet}; UMA through FAIR Chemistry at \url{https://github.com/facebookresearch/fairchem}; and NequIP-OAM-L through the NequIP model hub at \url{https://www.nequip.net/}.}

\section*{Benchmark reproducibility}
\label{ssec}

To support reproducibility of this Benchmark study, the input structures, molecular dynamics scripts, analysis notebooks, environment setup files, and model-loading workflows are provided at
\url{https://github.com/ziyuhuang652/Benchmark_FoundationModel_LunarRegolith}.
Because the six foundation models require different software dependencies, separate conda environments are used as documented in the repository. The reported trajectories, structural analyses, and performance benchmarks can be regenerated from the supplied scripts and input files.

%%============================================================
\bibliography{LunarFM}
\bibliographystyle{abbrv}

%% --- Foundation models ---

\end{document}